\documentclass[]{IAC_style}

\usepackage{amsmath}   
\usepackage{hyperref}  

\usepackage{tikz}
\usepackage[square,sort,comma,numbers]{natbib}
\usepackage{siunitx}
\usepackage{url}
\usepackage{graphicx} 
\usepackage{multirow}

\newcommand{\mb}[1]{\mathbf{#1}}

\DeclareMathOperator{\diag}{diag}

\DeclareMathOperator{\EX}{\mathbb{E}}

\usepackage{thmtools}

\usepackage{wrapfig}

\def\phi{\varphi}

\begin{document}

\IACpaperyear{2024} 
\IACpapernumber{IAC-24-C1-5-5-x90287} 
\IAClocation{Milan, Italy} 
\IACdate{14-18 October 2024} 

\IACcopyrightB{the International Astronautical Federation (IAF)}

\title{Certifying Guidance \& Control Networks: Uncertainty Propagation to an Event Manifold}

\IACauthor{Sebastien Origer$^{\orcidlink{0000-0003-1591-2099}}$}{2}{1}{0}
\IACauthor{Dario Izzo$^{\orcidlink{0000-0002-9846-8423}}$}{2}{0}{0}
\IACauthor{Giacomo Acciarini$^{\orcidlink{0000-0001-8707-8940}}$}{2}{0}{3}
\IACauthor{Francesco Biscani$^{}$}{2}{0}{4}
\IACauthor{Rita Mastroianni$^{\orcidlink{0000-0002-9211-3007}}$}{2}{0}{0}
\IACauthor{Max Bannach$^{\orcidlink{0000-0002-6475-5512}}$}{2}{0}{0}
\IACauthor{Harry Holt$^{\orcidlink{0000-0002-2035-2949}}$}{2}{0}{0}

\IACauthoraffiliation{Advanced Concepts Team, European Space Research and Technology Centre (ESTEC), Noordwijk, The Netherlands}
\IACauthoraffiliation{Surrey Space Centre, University of Surrey, Guildford, United Kingdom}
\IACauthoraffiliation{ESA/ESOC, Darmstadt, Germany}

\abstract{We perform uncertainty propagation on an event manifold for Guidance \& Control Networks (G\&CNETs), aiming to enhance the certification tools for neural networks in this field.
This work utilizes three previously solved optimal control problems with varying levels of dynamics nonlinearity and event manifold complexity. 
The G\&CNETs are trained to represent the optimal control policies of a time-optimal interplanetary transfer, a mass-optimal landing on an asteroid and energy-optimal drone racing, respectively.
For each of these problems, we describe analytically the terminal conditions on an event manifold with respect to initial state uncertainties. Crucially, this expansion does not depend on time but solely on the initial conditions of the system, thereby making it possible to study the robustness of the G\&CNET at any specific stage of a mission defined by the event manifold.
Once this analytical expression is found, we provide confidence bounds by applying the Cauchy-Hadamard theorem and perform uncertainty propagation using moment generating functions.
While Monte Carlo-based (MC) methods can yield the results we present, this work is driven by the recognition that MC simulations alone may be insufficient for future certification of neural networks in guidance and control applications.
}

\maketitle



\section{Introduction}
Guidance and Control Networks (G\&CNETs) are emerging as a promising type of neural network for enhancing onboard autonomy and seamlessly incorporating optimality principles into spacecraft \cite{sanchez2018real, izzo2021real, li2019neural, federici2021deep, hovell2020deep, cheng2018real, dario_seb_gcnet}. They provide an alternative to conventional model predictive control schemes (MPC) \cite{eren2017model} by leveraging advancements in machine learning.
Yet today, neural networks are still seen as inscrutable algorithms, sometimes referred to as "black boxes". 
With the increasing demand for spacecraft autonomy, it is necessary to develop methods that can assess the robustness of Guidance \& Control Networks, akin to stability analysis in control theory.
Simply evaluating the neural network over countless Monte Carlo simulations is not only time-consuming, it also does not provide a rigorous answer to the question: "Will my G\&CNET behave as intended when presented with a state it has never seen before?".
To address these limitations, past work~\cite{berz1988differential, berz1998verified, jorba2005software, jones2013nonlinear, park2006nonlinear} has already focused on the use of high-order Taylor expansions of the neural flow $\phi(\mathbf{x}_0,t)$ as a tool to gain insights into the uncertainty of such networks and potentially further tune their parameters.
The neural flow can be computed using methods such as Differential Algebra (DA), generalized dual numbers, or by integrating the variational equations.
Expanding the neural flow, however, limits the use cases to the ones where one is either interested in the state of the system after a specific amount of time has elapsed or for systems where an equilibrium point is likely to be reached after "enough" time has elapsed. Consider for instance the following requirement in the context of a neuro-controlled asteroid landing scenario:
\begin{quote}
The G\&C algorithm shall steer the spacecraft to an altitude of $1$ km $\pm 5$ m above the asteroid surface and ensure a relative velocity of $\leq 15$ m/s. The algorithm must achieve this relative velocity within the specified limit in at least 95\% of scenarios, given an initial state uncertainty of $\pm 5\%$ in spacecraft mass.
\end{quote}
Unless we revert back to a Monte Carlo-based approach, we cannot verify such a requirement by expanding the neural flow because the time at which the spacecraft will cross this boundary (defined as $1$ km in altitude above the asteroid surface) is not known in advance and it is not an equilibrium point. However we need to be able to comply with such requirements in order to increase the confidence in G\&CNETs and eventually deploy these on real space missions.
We show how requirements such as the one above can be addressed by cleverly manipulating the High Order Taylor Maps (HOTMs) of our systems and subsequently using two existing methods (Cauchy-Hadamard theorem and uncertainty propagation through moment generating functions \cite{acciarini2024mgf}):
\begin{enumerate}
    \item \textbf{Expansion on an event manifold}: Instead of expanding the neural flow, this paper proposes to do the expansion of the terminal conditions on an event manifold $\phi^*(\mathbf{x}_0,t)=\phi^*(\mathbf{x}_0,f(\mathbf{x}_0)) = \phi^*(\mathbf{x}_0)$. These tensors do not depend on time $t$ anymore but solely on the initial conditions $\mathbf{x}_0$. This can be done by inverting the HOTMs to analytically solve for the time at which the event is triggered. 
    The establishment of such analytical expressions, allows one to describe, for example, the terminal conditions on an event manifold acquired as a function of selected control parameters including initial conditions or other uncertainties, thereby providing insights into the robustness of the system around a nominal trajectory.
    \item \textbf{Confidence bounds}: The Cauchy-Hadamard theorem provides us with confidence bounds on the power series convergence.
    \item \textbf{Uncertainty propagation}: By staying within the bounds provided by the Cauchy-Hadamard theorem we can then confidently apply uncertainty propagation through moment generating functions \cite{acciarini2024mgf}.
\end{enumerate}

We show the versatility of this methodology by applying it to three test cases, each possessing different optimal control objectives, distinct dynamics/timescales, and event manifolds of varying complexity \cite{dario_seb_gcnet,origer2023guidance, origer2024closing}:

\begin{itemize}
    \item \textit{Interplanetary transfer}: 
    The goal of this problem is to learn the optimal thrust direction in order to perform a time-optimal, constant acceleration, low-thrust interplanetary transfer from the asteroid belt to a target circular orbit. The problem is characterized by its nonlinear dynamics and a typical optimal trajectory has a time-of-flight on the order of years \cite{dario_seb_gcnet}. The event manifold is the sphere of influence of the target planet, see Fig.\ref{fig:event_transfer}.
    \begin{figure}[h!]
       \centering
    \includegraphics[width=\columnwidth]{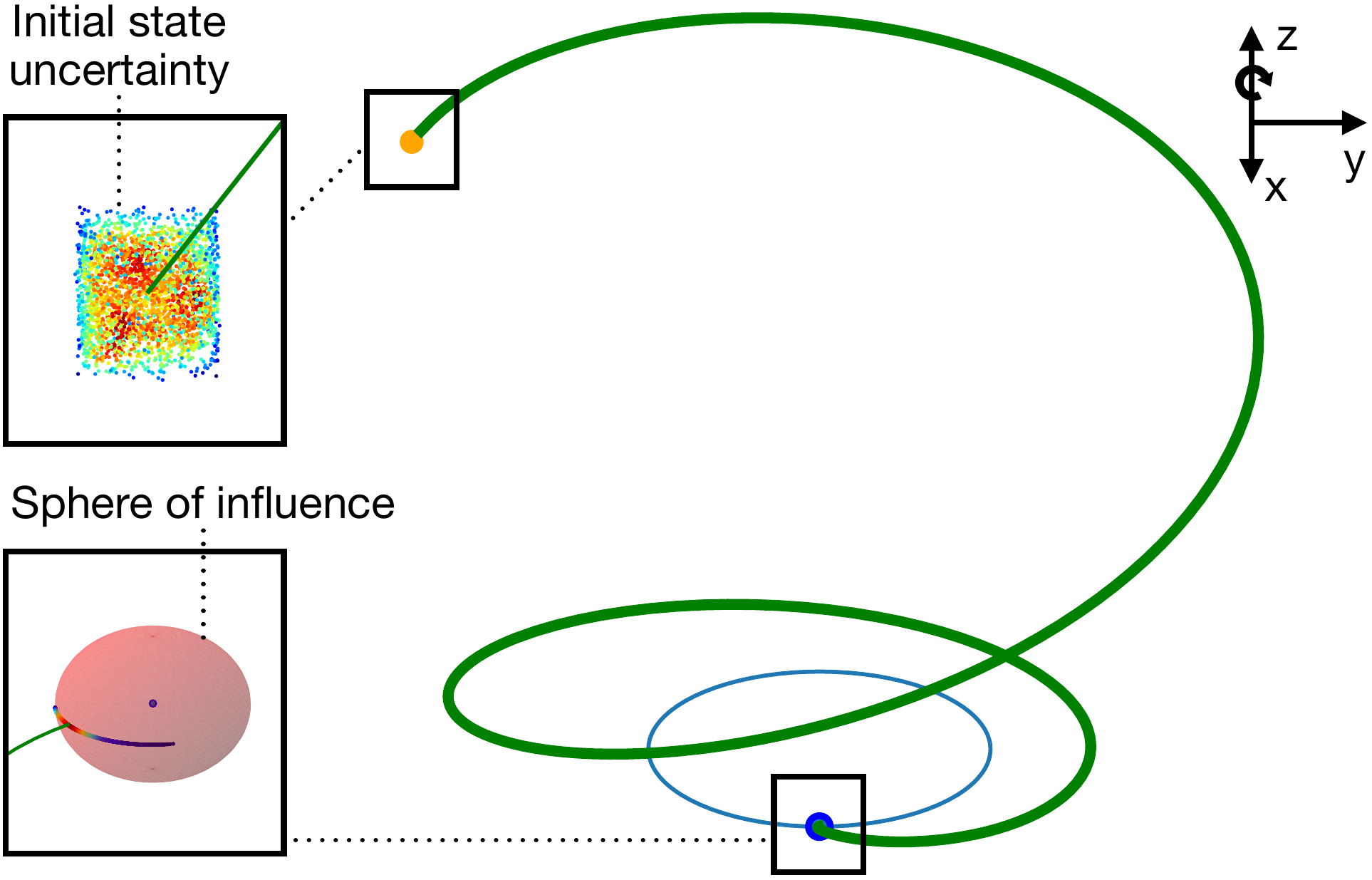}
    \caption{Interplanetary transfer: Uncertainty propagation of initial conditions onto the sphere of influence of the target planet. The scattered initial and final conditions are colored based on the local density of points (red and blue indicate high- and low-density regions, respectively).}\label{fig:event_transfer}
    \end{figure}
    \item \textit{Asteroid landing}: 
    The G\&CNET for this problem learns the optimal thrust direction and a
    discontinuous `bang-bang' profile for the throttle to perform a mass-optimal landing on the asteroid Psyche. 
    The equations of motion are nonlinear and a typical optimal trajectory has a time-of-flight on the order of minutes to hours \cite{origer2024closing}. 
    The event manifold is a complex three-dimensional shape representing the boundary at $1$ km altitude above the asteroid's surface, see Fig.\ref{fig:event_landing}.
    \begin{figure}[h!]
       \centering
    \includegraphics[width=\columnwidth]{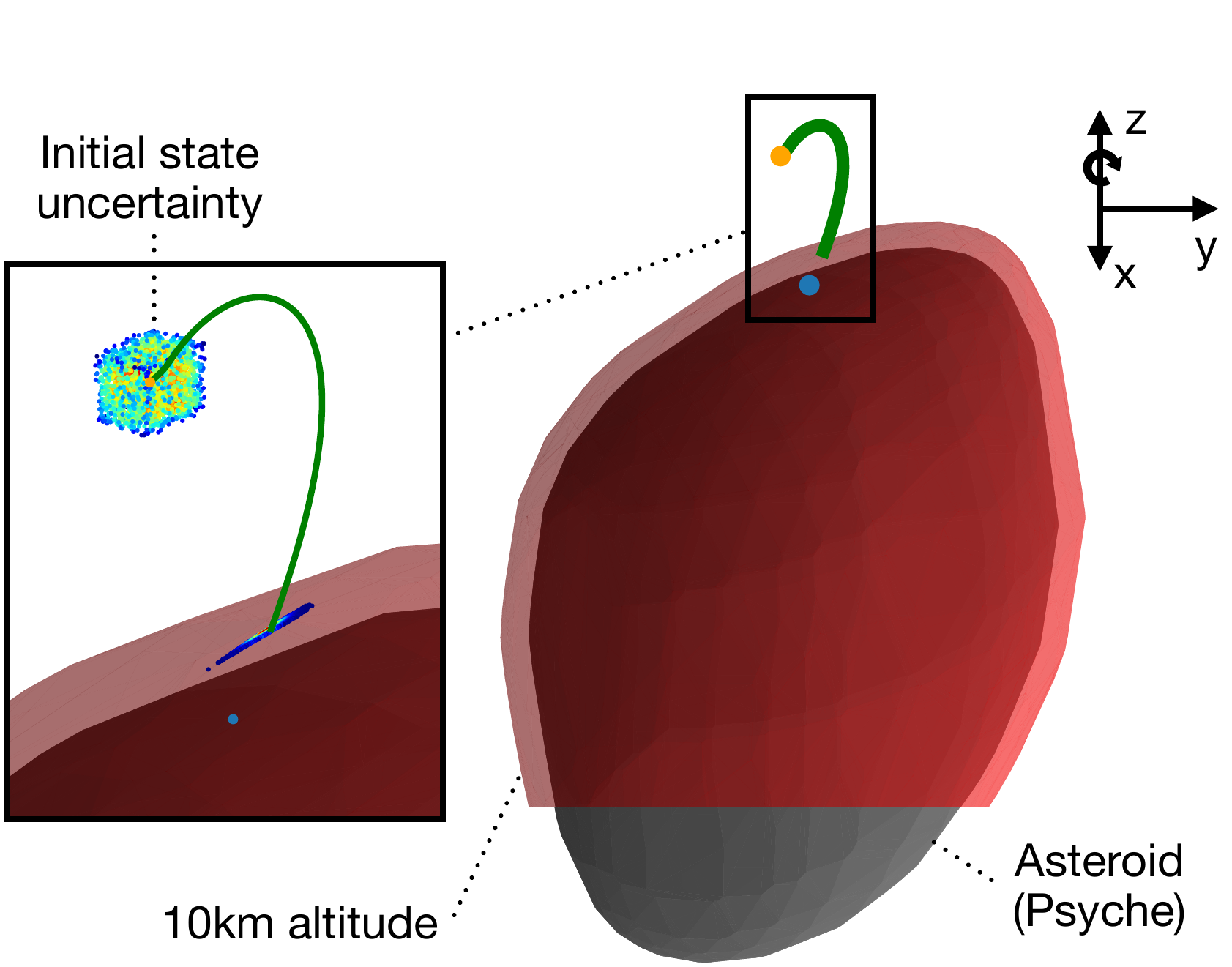}
    \caption{Asteroid landing: Uncertainty propagation of initial conditions onto the complex shape describing a constant altitude from the asteroid surface. We show the manifold at $10$km altitude to make it more visible, but use an altitude of $1$km in the rest of this work. The scattered initial and final conditions are colored based on the local density of points (red and blue indicate high- and low-density regions, respectively).}\label{fig:event_landing}
    \end{figure}
    \item \textit{Drone racing}: This is an energy-optimal control problem that aims to steer a quadcopter from a range of possible initial conditions through a square gate. This problem possesses highly nonlinear dynamics and a typical optimal trajectory time-of-flight is on the order of seconds \cite{origer2023guidance}. The event manifold is a simple two-dimensional square gate, see Fig.\ref{fig:event_drone}.
    \begin{figure}[h!]
       \centering
    \includegraphics[width=\columnwidth]{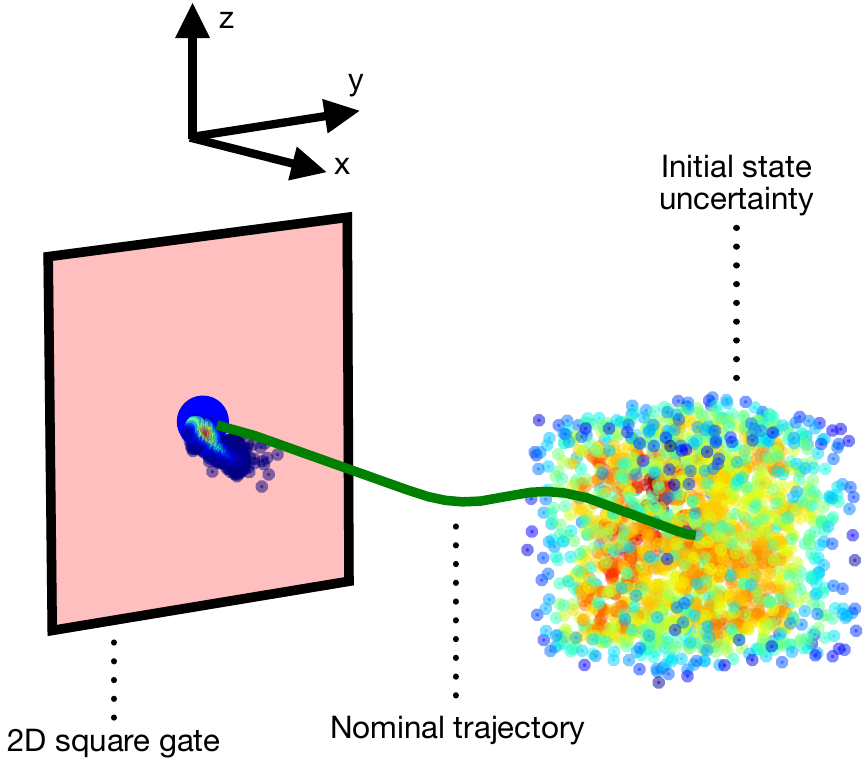}
    \caption{Drone racing: Uncertainty propagation of initial conditions onto a two-dimensional square gate. The scattered initial and final conditions are colored based on the local density of points (red and blue indicate high- and low-density regions, respectively).}\label{fig:event_drone}
    \end{figure}
\end{itemize}

\section{Methods}

\subsection{High-Order Taylor Maps}

We consider a dynamical system described by a set of ordinary differential equations:
\begin{equation}
    \dot{\pmb{x}}=\pmb{f}(\pmb{x},\pmb{q})
\end{equation}
where $\pmb{f}\in \mathbb{R}^n$ is the vector describing the dynamics, $\pmb{x}\in \mathbb{R}^n$ is the state vector and $\pmb{q}\in \mathbb{R}^m$ a vector of parameters.

By Taylor expanding of the flow $\pmb{x}_f(t;\pmb{x}_0,\pmb{q})$, we obtain:
\begin{equation}
    \delta x^i_f= \mathcal{P}_{i}^k(\delta \pmb{x}_0, \delta \pmb{q})+\mathcal{O}(k)
    \label{eq:taylor_expansion}
\text{,}
\end{equation}

with $\delta x_f^i$ being the $i$-th component of the state deviation at the final time and $\mathcal{P}_{i}^k$ the $k$-the order Taylor polynomial for the $i$-th component of the final state deviations. The symbol $\mathcal{O}(k)$ indicates the terms of the series that have an order higher than $k$. 
Also, we use $\delta \pmb{x}_0$ $\delta \pmb{q}$ to refer to perturbations w.r.t. the nominal initial state $\overline{\pmb{x}}_0$, and parameters $\overline{\pmb{q}}$. 

By grouping the initial state perturbations and parameters deviation into the same vector: $\delta \pmb{z}=[\delta\pmb{x}_0^T, \delta \pmb{q}^T]^T$, and using multi-index notation, we write Eq.~\eqref{eq:taylor_expansion} in a compact form as:
\begin{equation}
    \delta x_f^i\approx \mathcal{P}_i^k(\delta \pmb{z})=\sum_{|\alpha|=1}^k \dfrac{1}{\alpha!}(\partial^\alpha x_{f}^i)\bigg|_{(\overline{\pmb{x}}_0, \overline{\pmb{q}})}\delta \pmb{z}^\alpha\text{,}
\label{eq:taylor_polynomials_multi_index}
\end{equation}
where $\alpha=(\alpha_1,\dots \alpha_n)$ is the $n$-tuple of non-negative integers. Each $\alpha_i$ is associated with the $i$-th component of the vector $\pmb{x}_f$. 
The factorial of $\alpha$ is defined as $\alpha!=\alpha_1!\dots\alpha_n!$, while $|\alpha|=\sum_{j=0}^M\alpha_j$ must be taken over all possible combinations of $\alpha_j \in \mathbb{N}$. The taylor coefficients of the series in Eq.~\eqref{eq:taylor_polynomials_multi_index} are also referred to as state transition tensors.

\subsection{Taylor expansion on an event manifold}
A similar concept to state transition tensors can also be defined in the case in which the integration has to be stopped at an event manifold: in this scenario, we refer to the corresponding Taylor coefficients as event transition tensors (ETT)~\cite{acciarini2024mgf}. For these cases, we use the concept of the event manifold, that is, an implicitly defined equation such that:
\begin{equation}
e(\pmb{x}_f(t^*;\pmb{x}_0,\pmb{q}),\pmb{q})=0\text{,}
\label{eq:event_equation}
\end{equation}
with $t^*$ defined as the event trigger time. 

By including the integration time as a parameter in the Taylor expansion of the flow shown in Eq.~\eqref{eq:taylor_polynomials_multi_index}, and stopping the integration of the nominal state at the event crossing, one can partially invert the map~\cite{hawkes1999modern, armellin2021collision}, and find the Taylor polynomial associated with the trigger time. Then, by imposing the event satisfaction through Eq.~\eqref{eq:event_equation}, one ensures that any perturbation around the nominal initial state and parameters still satisfies the event equation (provided that the perturbations are within the radius of convergence of the Taylor series). Finally, substituting back the trigger time polynomial that satisfies the event equation inside the expansion of the flow shown in Eq.~\eqref{eq:taylor_polynomials_multi_index}, one obtains the Taylor map that directly describes the state at the event crossing. Given that this is still a polynomial, all the techniques that we will discuss in Sec.~\ref{sec:radius_of_convergence_of_multivariate_taylor_series} and \ref{sec:up_through_mgf} will still hold.

For a tutorial on a simple map inversion problem, we refer the authors to the \textit{heyoka}\footnote{Map inversion tutorial with \textit{heyoka}: \url{ https://bluescarni.github.io/heyoka.py/notebooks/map_inversion.html}} python library~\cite{biscani2021revisiting}.
 




\subsection{Radius of Convergence of Multivariate Taylor Series}
\label{sec:radius_of_convergence_of_multivariate_taylor_series}

Results from calculus provide a way to compute the radius of convergence of infinite power series like the Taylor series. The ratio test~\cite{bromwich2005introduction, izzo2020stability} and the Cauchy-Hadamard theorem~\cite{cauchy1821cours,hadamard1892essai} are two of the most used criteria for establishing whether a series converges and what is the convergence radius. Nonetheless, expanding a Taylor series indefinitely is often not practical, and the series is truncated at a given order: in such scenarios, these tests only provide an approximation of the convergence radius.
Clearly, if the series converges, the two criteria must converge to the same radius when expanded at infinity. Monitoring how the radius of convergence evolves as the truncation order is increased can provide insights into which criterion is most suited for the given application. As observed also in \cite{acciarini2024mgf}, in all our experiments we find that the ratio test has a more oscillating behavior for relatively low truncation orders, thereby making it more difficult to reliably estimate the convergence radius. In contrast, the Cauchy-Hadamard test displayed a more stable behavior and we thereby opted for that.

We write the Cauchy-Hadamard convergence radius for multivariate Taylor series as~\cite{bekbaev2013radius}:
\begin{equation}
R_{c}=1/\lim_{k\rightarrow\infty}\textrm{sup}_{|\alpha|=k}|a_{\alpha}\big( \alpha!/k!\big)^{1/2}|^{1/k}
\text{,}
    \label{eq:cauchy_hadamard}
\end{equation}
where $a_{\alpha}$ is the same as the one defined in Eq.~\eqref{eq:taylor_polynomials_multi_index}.

\subsection{Uncertainty propagation through moment generating function}
\label{sec:up_through_mgf}
In this work, we use the results from \cite{acciarini2024mgf}, which extend the work of Park \& Scheeres to non-Gaussian scenarios and event manifolds~\cite{park2006nonlinear}, to propagate moments of uniform probability density function at future times and/or events. We hereby work with initial distributions that are uniformly distributed, and in which each component is (initially) independent. The probability density function (pdf) for each component of the random vector can be written as:

\begin{equation}
p(x_i)=\dfrac{1}{b_i-a_i}, \ a_i\le x_i \le b_i\text{,}
\end{equation}
with $a_i$ and $b_i$ being the lower and upper bounds of each random variable. For a uniform pdf, the moment-generating function (mgf) can be written as:
\begin{equation}
    M_{x_i}(t)=\dfrac{e^{tb_i}-e^{ta_i}}{t(b_i-a_i)}\text{.}
\end{equation}

Then, as a consequence of the fact that the random variables are initially independently distributed, the mgf for the $n$-dimensional random vector can be written as product of the mgf of each variable:
\begin{equation}
    M_{\pmb{x}}(\pmb{t})=\prod_{i=1}^n M_{x_i}(t_i)\text{.}
\end{equation}

As shown in \cite{acciarini2024mgf}, we can then take the expectation of the state at the final time or event, and leverage the linearity of the expectation operator, and the connection between mgf and expectations, to obtain a method to propagate moments of the distribution.
For instance, for the first moment (i.e., the mean) of the pdf, by taking the expectation of perturbations around the nominal solution flow, we get:

\begin{equation}
    \EX[\delta x_f^i]\approx \EX[\mathcal{P}_i^k(\delta \pmb{z})]=\sum_{|\alpha|=1}^k \dfrac{1}{\alpha!}(\partial^\alpha x_f^i)\bigg|_{(\overline{\pmb{x}}_0, \overline{\pmb{q}})}\EX[\delta \pmb{z}^\alpha]\text{.}
    \label{eq:expectation_flow_perturbations}
\end{equation}

Similarly, the second and higher moments expression can also be derived~\cite{acciarini2024mgf}.
Note that this methodology is not limited to uniform distributions, but to any distribution for which the moment-generating function exists.

\subsection{Optimal Control Problems}

\subsubsection*{Interplanetary transfer}
The target body is in a circular orbit of radius $R$ \cite{dario_seb_gcnet,origer2024closing}.
We introduce a rotating frame $\mathcal F = [\hat{\mathbf i}, \hat{\mathbf j}, \hat{\mathbf k}]$ of angular velocity $\boldsymbol \Omega= \Omega \mathbf{\hat{k}}= \sqrt{\mu/R^3} \hat {\mathbf k}$.
Thus, the position of the target body $R\hat{\mathbf i}$ remains stationary in $\mathcal F$.
The equations of motion are:
\begin{equation}
\label{eq:dyn_transfer}
\left\{ 
\begin{array}{l}
    \dot{x} = v_x  \\
    \dot{y} = v_y \\
    \dot{z} = v_z \\
    \dot{v}_x = -\frac{\mu}{r^3}x + 2 \Omega v_y + \Omega^2 x +\Gamma i_x \\
    \dot{v}_y = -\frac{\mu}{r^3}y - 2 \Omega v_x + \Omega^2 y +\Gamma i_y \\
    \dot{v}_z = -\frac{\mu}{r^3}z +\Gamma i_z
\end{array}
\right.\, .
\end{equation}

The state vector $\mathbf{x}$ consists of the position $\mathbf{r}=[x,y,z]$ and velocity $\mathbf{v}=[v_x,v_y,v_z]$, both expressed in the rotating frame $\mathcal{F}$.
Here $r = \sqrt{x^2+y^2+z^2}$ and $\mu$ denotes the gravitational constant of the Sun.
The system is controlled by the thrust direction, represented by the unit vector $\mathbf{\hat{t}} = [i_x, i_y, i_z]$, which produces an acceleration of magnitude $\Gamma = 0.1$ mm/s$^2$.
The goal of this time-optimal control problem is to determine a (piece-wise continuous) function for $\mathbf{\hat{t}}(t)$ and the optimal time-of-flight $t_f$, where $t \in [t_0,t_f]$, so that, following the dynamics described by Eq.\ref{eq:dyn_transfer}, the state is steered from any initial state $\mathbf{r}_0$, $\mathbf{v}_0$ to the desired target state $\mathbf{r}_t = R\mathbf{\hat{i}}$, $\mathbf{v}_t = \mathbf{0}$.
Since the goal is to be time-optimal, we minimize the following cost function:
$J = t_f-t_0$ \cite{dario_seb_gcnet}. 
We solve this optimal control problem with an indirect method \cite{pontryagin}. The reader is referred to \cite{origer2024closing} for a detailed derivation, which is omitted here for brevity.
While we train the G\&CNET to steer the spacecraft all the way to the target planet, in this work we are only interested in the behaviour of the network up to the sphere of influence of the target planet.
This is motivated by the fact that in a real world scenario, it is rather unlikely that a single network would take care of guiding and controlling the spacecraft for the entire mission. In this case, after roughly $4.6$ years of low-thrust transfer, a different G\&C scheme would likely intervene once the sphere of influence is reached, correcting accumulated errors in position and velocity with a couple of high thrust impulses.
Using the following event function $e(x,y,z)$, where $R_{\text{SOI}}= 924,000$ km is the radius of the sphere of influence:
\begin{equation}
\label{eq:event_transfer}
    e(x,y,z) = (x - R)^2 + y^2 + z^2 - R_{\text{SOI}}^2
\end{equation}
we can define the terminal event manifold as the points where $e=0$.

\subsubsection*{Asteroid landing}\label{subsec:ast}
We introduce the rotating frame $\mathcal R = [\hat{\mathbf i}, \hat{\mathbf j}, \hat{\mathbf k}]$ of angular velocity $\boldsymbol \omega \hat {\mathbf k}$, such that the asteroid remains stationary within $\mathcal R$ \cite{origer2024closing}.
The equations of motion are:
\begin{equation}
\label{eq:dyn_landing}
\left\{ 
\begin{array}{l}
    \dot{x} = v_x \\
    \dot{y} = v_y \\
    \dot{z} = v_z \\
    \dot{v}_x = -\frac{\mu}{r^3}x + 2 \omega v_y + \omega^2 x +u\frac{c_1}{m} i_x \\
    \dot{v}_y = -\frac{\mu}{r^3}y - 2 \omega v_x + \omega^2 y +u\frac{c_1}{m} i_y \\
    \dot{v}_z = -\frac{\mu}{r^3}z +u\frac{c_1}{m} i_z \\
    \dot{m} = -u\frac{c_1}{I_{sp}g_0}
\end{array}
\right. \, .
\end{equation}
The state vector $\mathbf{x}$ consists of the position $\mathbf{r}=[x,y,z]$, velocity $\mathbf{v}=[v_x,v_y,v_z]$ (both expressed in the rotating frame $\mathcal R$), and mass $m$ of the spacecraft.
Here $r = \sqrt{x^2+y^2+z^2}$ and $\mu$ denotes the gravitational constant of Psyche.
The system is controlled by the thrust direction specified by the unit vector $\mathbf{\hat{t}} = [i_x, i_y, i_z]$ and the throttle $u \in [0,1]$. Model parameters are listed in App.\ref{app:0}.

The goal of this mass-optimal control problem with free final time $t_f$ and free final mass $m_f$ is to determine piece-wise continuous functions for $u(t)$ and $\mathbf{\hat{t}}(t)$, where $t\in [t_0,t_f]$, so that, following the dynamics described by Eq.\ref{eq:dyn_landing}, the state of the spacecraft is steered from any initial state $\mathbf{r}_0$, $\mathbf{v}_0$, $m_0$ to the desired target state $\mathbf{r}_t$, $\mathbf{v}_t$. 
Since the goal is to be mass-optimal, we minimize the following cost function: $J = m_0-m_f$. 
This problem can be solved using an indirect method (Pontryagin's Maximum Principle \cite{pontryagin}); the details are omitted here for brevity but can be looked up in \cite{origer2024closing}. 
Here, again, we train the G\&CNET to steer the spacecraft all the way to the target landing site (pinpoint landing), but we are only interested in analyzing the robustness of the G\&CNET up to a certain altitude above the asteroid surface. The event equation in this case is nontrivial, given the complex three-dimensional shape of the asteroid.
We solve this problem by training a small feedforward neural network $\mathcal{E}(x,y,z)$, with only 56 parameters, to approximate this boundary ($1$ km altitude). The network takes as input the current position $\mathbf{r}=[x,y,z]$ and outputs a positive value for points above $1$ km altitude, zero at $1$ km, and a negative value for points below $1$ km. We use the SIREN architecture \cite{sitzmann2019siren}, which has shown to accurately represent complex shapes implicitly \cite{Izzo2022Geodesy, acciarini2024eclipsenet}. Using triangular meshes which describe the shape of the asteroid and the Möller–Trumbore intersection algorithm, one can efficiently determine whether a random point is inside or outside of the desired boundary. Training the neural network then becomes a simple regression task on millions of randomly generated points using behavioral cloning.
Note that the event machinery in the \textit{heyoka} python library \cite{biscani2021revisiting} allows for a seamless integration of feedforward neural networks as the event equation. The terminal event manifold is then defined as the points where $\mathcal{E}=0$.

\subsubsection*{Drone racing}
We utilize two coordinate frames (see App.\ref{app:1}), one inertial world frame $\mathcal{W}$ centered in the middle of gate through which the drone will fly through and one body frame $\mathcal{B}$ attached to the drone. The drone model has 16 states $\mb{x} = [\mb{p}, \mb{v}, \boldsymbol{\lambda}, \boldsymbol\Omega, \boldsymbol{\omega}]$ and four control inputs $\mb{u} = [u_1, u_2, u_3, u_4]$ \cite{origer2023guidance}.
The state vector $\mb{x}$ consists of the position $\mb{p}=[x,y,z]$ and velocity $\mb{v}=[v_x,v_y,v_z]$, both expressed in the world frame.
The Euler angles $\boldsymbol{\lambda}=[\phi, \theta, \psi]$, which determine the orientation of the body frame, the angular velocities $\boldsymbol{\Omega}=[p,q,r]$ within the body frame, and the angular propeller rates $\boldsymbol{\omega}=[ \omega_1, \omega_2, \omega_3, \omega_4]$.
The control inputs $\mb{u} = [u_1, u_2, u_3, u_4]$ are restricted within $u_i \in [0,1]$, where $u_i=0$ corresponds to the minimum ($\omega_{min}$) angular rate and $u_i=1$  to the maximum ($\omega_{max}$) angular rate.
The equations of motion are:
\begin{equation}
\label{eq:dyn}
f(\mb{x}, \mb{u}) = \left\{ 
\begin{array}{l}
        \dot{\mb{p}} = \mb{v} \\
        \dot{\mb{v}} = \mb{g} + R(\boldsymbol{\lambda}) \mb{F} \\
        \dot{\boldsymbol\lambda} = Q(\boldsymbol{\lambda}) \mb{\Omega} \\
        I \dot{\boldsymbol{\Omega}} = - \boldsymbol{\Omega} \times I \boldsymbol{\Omega} + \mb{M}\\
        \dot{\boldsymbol\omega}  = ((\omega_{max}-\omega_{min}) \mb{u} +\omega_{min} - \boldsymbol\omega)/\tau
\end{array}
\right.  ,
\end{equation}
where $I=\diag (I_x, I_y, I_z)$ is the moment of inertia matrix and $\mb{g} = [0, 0, g]^T$ with $g=9.81$~\si{\meter\per\second\squared} is the gravitational acceleration. 
The rotation matrix $R(\mb{\boldsymbol\lambda})$, forces $\mb{F}$, moments $\mb{M}$, and model coefficients are detailed in App.\ref{app:1}.
This optimal control problem aims to minimize the total energy consumption, using as cost function $J(\mathbf{u}, T) = \int_{0}^{T} ||\mb{u}(t)||^2 dt$.
Let $X$ represent the state space and $U$ the set of allowable controls. The goal of the optimal control problem is to identify the optimal control policy $\mb{u} : [0,1] \rightarrow U$, which steers the quadcopter from some initial conditions $\mb{x}_0$ to a specified set of target conditions $S$, while minimizing the cost function $J(\mb{u}, T)$.
To alleviate numerical instabilities and simplify this problem, we set the event equation $e(x)$ as the plane $x=0$ (on which the square gate lies) and discard trajectories in the final analysis which missed the square gate. Hence the terminal event manifold is the plane $x=0$.

\subsection{Behavioral cloning}
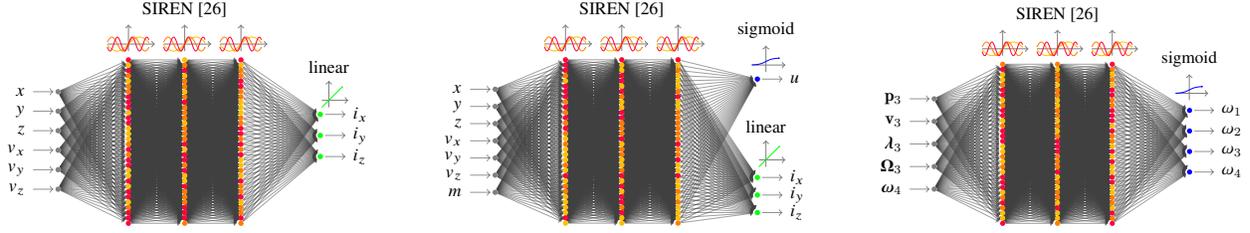
\begin{figure*}[h]
  \centering
    \resizebox{0.32\textwidth}{!}{
\def\layersep{2cm}
\def\nodesep{0.3cm}
\def\linewidthact{0.18mm}
\definecolor{americanrose}{rgb}{1.0,0.01,0.24}
\definecolor{or}{rgb}{1.0,0.49,0.0}
\definecolor{yell}{rgb}{1.0,0.75,0.0}

\newcommand{\randomcolor}{
    \pgfmathsetmacro{\randnum}{int(random(1,3))}
    \ifnum\randnum=1
        \def\colorselect{americanrose}
    \else\ifnum\randnum=2
        \def\colorselect{or}
    \else
        \def\colorselect{yell}
    \fi\fi
}
\newcommand{\hiddenneuronrandomcolor}{
    \randomcolor
    \tikzstyle{hidden neuron softplus}=[neuron_hidden, fill=\colorselect]
}

\begin{tikzpicture}[shorten >=1pt,->,draw=black!50, node distance=\layersep, scale=0.7]
    \tikzstyle{every pin edge}=[<-,shorten <=1pt,line width=0.5pt,color=black!85, opacity=0.5]

    \tikzset{mystyle/.style={line width=0.025mm,color=black!75, opacity=0.5}}

    \tikzstyle{neuron}=[circle,fill=black!25,minimum size=3pt,inner sep=0pt]

    \tikzstyle{neuron_hidden}=[circle,fill=black!25,minimum size=3pt,inner sep=0pt]
    
    \tikzstyle{input neuron}=[neuron, fill=gray];
    \tikzstyle{output neuron}=[neuron, fill=blue];
    \tikzstyle{output neuron2}=[neuron, fill=green];

    \tikzstyle{hidden neuron softplus}=[neuron_hidden, fill=blue!50];

    \tikzstyle{annot} = [text width=4em, text centered]

    \node[input neuron, pin=left:$x$] (I-1) at (0,-\nodesep*1*1.85) {};
    \node[input neuron, pin=left:$y$] (I-2) at (0,-\nodesep*2*1.85) {};
    \node[input neuron, pin=left:$z$] (I-3) at (0,-\nodesep*3*1.85) {};
    \node[input neuron, pin=left:$v_x$] (I-4) at (0,-\nodesep*4*1.85) {};
    \node[input neuron, pin=left:$v_y$] (I-5) at (0,-\nodesep*5*1.85) {};
    \node[input neuron, pin=left:$v_z$] (I-6) at (0,-\nodesep*6*1.85) {};


    \foreach \name / \y in {1,...,32}
        \hiddenneuronrandomcolor
        \path[yshift=0.5cm]
            node[hidden neuron softplus] (H1-\name) at (\layersep,-\nodesep*0.5*\y) {};

    \foreach \name / \y in {1,...,32}
        \hiddenneuronrandomcolor
        \path[yshift=0.5cm]
            node[hidden neuron softplus] (H2-\name) at (1.8*\layersep,-\nodesep*0.5*\y) {};
            
    \foreach \name / \y in {1,...,32}
        \hiddenneuronrandomcolor
        \path[yshift=0.5cm]
            node[hidden neuron softplus] (H3-\name) at (2.6*\layersep,-\nodesep*0.5*\y) {};

    \node[output neuron2, pin={[pin edge={->}]right:$i_x$}, right of=H3-3] (O-1) at (2.3*\layersep,-1.2)  {};
    \node[output neuron2, pin={[pin edge={->}]right:$i_y$}, right of=H3-3] (O-2) at (2.3*\layersep,-1.8)  {};
    \node[output neuron2, pin={[pin edge={->}]right:$i_z$}, right of=H3-3] (O-3) at (2.3*\layersep,-2.4)  {};

    \pgfdeclarelayer{background}
    \pgfsetlayers{background,main}

    \begin{pgfonlayer}{background}

    \foreach \source in {1,...,6}
        \foreach \dest in {1,...,32}
            \path[mystyle] (I-\source) edge (H1-\dest);
            
    \foreach \source in {1,...,32}
        \foreach \dest in {1,...,32}
            \path[mystyle] (H1-\source) edge (H2-\dest);


    \foreach \source in {1,...,32}
        \foreach \dest in {1,...,32}
            \path[mystyle] (H2-\source) edge (H3-\dest);

    \foreach \source in {1,...,32}
        \foreach \dest in {1,...,3}
            \path[mystyle] (H3-\source) edge (O-\dest);

    \end{pgfonlayer}

    \pgfmathsetmacro{\scale}{0.2}
    \pgfmathsetmacro{\Ox}{10}
    \pgfmathsetmacro{\Oy}{4}

    \draw[scale=\scale, shift = {(\Ox, \Oy)}] (-3,0) -- (4,0) node[right] {};
    \draw[scale=\scale, shift = {(\Ox, \Oy)}] (0,-1) -- (0,3) node[above] {};
    \draw[-, scale=\scale,domain=-3:3,smooth,variable=\x,yell, shift = {(\Ox, \Oy)}, line width=\linewidthact ] 
        plot ({\x},{sin(1 * \x r)});
    \draw[-, scale=\scale,domain=-3:3,smooth,variable=\x,or, shift = {(\Ox, \Oy)}, line width=\linewidthact ] 
        plot ({\x},{sin(2 * \x r)});
    \draw[-, scale=\scale,domain=-3:3,smooth,variable=\x,americanrose, shift = {(\Ox, \Oy)}, line width=\linewidthact ] 
        plot ({\x},{sin(3 * \x r)});
    
    \draw[scale=\scale, shift = {(\Ox+8, \Oy)}] (-3,0) -- (4,0) node[right] {};
    \draw[scale=\scale, shift = {(\Ox+8, \Oy)}] (0,-1) -- (0,3) node[above] {SIREN \cite{sitzmann2019siren}};
    \draw[-, scale=\scale,domain=-3:3,smooth,variable=\x,yell, shift = {(\Ox+8, \Oy)}, line width=\linewidthact ] 
        plot ({\x},{sin(1 * \x r)});
    \draw[-, scale=\scale,domain=-3:3,smooth,variable=\x,or, shift = {(\Ox+8, \Oy)}, line width=\linewidthact ] 
        plot ({\x},{sin(2 * \x r)});
    \draw[-, scale=\scale,domain=-3:3,smooth,variable=\x,americanrose, shift = {(\Ox+8, \Oy)}, line width=\linewidthact ] 
        plot ({\x},{sin(3 * \x r)});
    
    \draw[scale=\scale, shift = {(\Ox+16, \Oy)}] (-3,0) -- (4,0) node[right] {};
    \draw[scale=\scale, shift = {(\Ox+16, \Oy)}] (0,-1) -- (0,3) node[above] {};
    \draw[-, scale=\scale,domain=-3:3,smooth,variable=\x,yell, shift = {(\Ox+16, \Oy)}, line width=\linewidthact ] 
        plot ({\x},{sin(1 * \x r)});
    \draw[-, scale=\scale,domain=-3:3,smooth,variable=\x,or, shift = {(\Ox+16, \Oy)}, line width=\linewidthact ] 
        plot ({\x},{sin(2 * \x r)});
    \draw[-, scale=\scale,domain=-3:3,smooth,variable=\x,americanrose, shift = {(\Ox+16, \Oy)}, line width=\linewidthact ] 
        plot ({\x},{sin(3 * \x r)});
    

    
    \draw[scale=\scale, shift = {(\Ox+28.5, \Oy-8)}] (-1,0) -- (3,0) node[right] {};
    \draw[scale=\scale, shift = {(\Ox+28.5, \Oy-8)}] (0,-1) -- (0,3) node[above] {linear};
    \draw[-, scale=\scale,domain=-1:2,smooth,variable=\x,green, shift = {(\Ox+28.5, \Oy-8)}, line width=\linewidthact ] plot ({\x},{(\x)});
    
\end{tikzpicture}


}
    \hfill
    \resizebox{0.32\textwidth}{!}{
\def\layersep{2cm}
\def\nodesep{0.3cm}
\def\linewidthact{0.18mm}
\definecolor{americanrose}{rgb}{1.0,0.01,0.24}
\definecolor{or}{rgb}{1.0,0.49,0.0}
\definecolor{yell}{rgb}{1.0,0.75,0.0}

\newcommand{\randomcolor}{
    \pgfmathsetmacro{\randnum}{int(random(1,3))}
    \ifnum\randnum=1
        \def\colorselect{americanrose}
    \else\ifnum\randnum=2
        \def\colorselect{or}
    \else
        \def\colorselect{yell}
    \fi\fi
}
\newcommand{\hiddenneuronrandomcolor}{
    \randomcolor
    \tikzstyle{hidden neuron softplus}=[neuron_hidden, fill=\colorselect]
}

\begin{tikzpicture}[shorten >=1pt,->,draw=black!50, node distance=\layersep, scale=0.7]
    \tikzstyle{every pin edge}=[<-,shorten <=1pt,line width=0.5pt,color=black!85, opacity=0.5]

    \tikzset{mystyle/.style={line width=0.025mm,color=black!75, opacity=0.5}}

    \tikzstyle{neuron}=[circle,fill=black!25,minimum size=3pt,inner sep=0pt]

    \tikzstyle{neuron_hidden}=[circle,fill=black!25,minimum size=3pt,inner sep=0pt]
    
    \tikzstyle{input neuron}=[neuron, fill=gray];
    \tikzstyle{output neuron}=[neuron, fill=blue];
    \tikzstyle{output neuron2}=[neuron, fill=green];

    \tikzstyle{hidden neuron softplus}=[neuron_hidden, fill=blue!50];

    \tikzstyle{annot} = [text width=4em, text centered]

    \node[input neuron, pin=left:$x$] (I-1) at (0,-\nodesep*1*1.63) {};
    \node[input neuron, pin=left:$y$] (I-2) at (0,-\nodesep*2*1.63) {};
    \node[input neuron, pin=left:$z$] (I-3) at (0,-\nodesep*3*1.63) {};
    \node[input neuron, pin=left:$v_x$] (I-4) at (0,-\nodesep*4*1.63) {};
    \node[input neuron, pin=left:$v_y$] (I-5) at (0,-\nodesep*5*1.63) {};
    \node[input neuron, pin=left:$v_z$] (I-6) at (0,-\nodesep*6*1.63) {};
    \node[input neuron, pin=left:$m$] (I-7) at (0,-\nodesep*7*1.63) {};


    \foreach \name / \y in {1,...,32}
        \hiddenneuronrandomcolor
        \path[yshift=0.5cm]
            node[hidden neuron softplus] (H1-\name) at (\layersep,-\nodesep*0.5*\y) {};

    \foreach \name / \y in {1,...,32}
        \hiddenneuronrandomcolor
        \path[yshift=0.5cm]
            node[hidden neuron softplus] (H2-\name) at (1.8*\layersep,-\nodesep*0.5*\y) {};
            
    \foreach \name / \y in {1,...,32}
        \hiddenneuronrandomcolor
        \path[yshift=0.5cm]
            node[hidden neuron softplus] (H3-\name) at (2.6*\layersep,-\nodesep*0.5*\y) {};

    \node[output neuron, pin={[pin edge={->}]right:$u$}, right of=H3-3] (O-1) at (2.3*\layersep,-0.2)  {};
    \node[output neuron2, pin={[pin edge={->}]right:$i_x$}, right of=H3-3] (O-2) at (2.3*\layersep,-3)  {};
    \node[output neuron2, pin={[pin edge={->}]right:$i_y$}, right of=H3-3] (O-3) at (2.3*\layersep,-3.5)  {};
    \node[output neuron2, pin={[pin edge={->}]right:$i_z$}, right of=H3-3] (O-4) at (2.3*\layersep,-4.0)  {};

    \pgfdeclarelayer{background}
    \pgfsetlayers{background,main}

    \begin{pgfonlayer}{background}

    \foreach \source in {1,...,7}
        \foreach \dest in {1,...,32}
            \path[mystyle] (I-\source) edge (H1-\dest);
            
    \foreach \source in {1,...,32}
        \foreach \dest in {1,...,32}
            \path[mystyle] (H1-\source) edge (H2-\dest);


    \foreach \source in {1,...,32}
        \foreach \dest in {1,...,32}
            \path[mystyle] (H2-\source) edge (H3-\dest);

    \foreach \source in {1,...,32}
        \foreach \dest in {1,...,4}
            \path[mystyle] (H3-\source) edge (O-\dest);

    \end{pgfonlayer}

    \pgfmathsetmacro{\scale}{0.2}
    \pgfmathsetmacro{\Ox}{10}
    \pgfmathsetmacro{\Oy}{4}

    \draw[scale=\scale, shift = {(\Ox, \Oy)}] (-3,0) -- (4,0) node[right] {};
    \draw[scale=\scale, shift = {(\Ox, \Oy)}] (0,-1) -- (0,3) node[above] {};
    \draw[-, scale=\scale,domain=-3:3,smooth,variable=\x,yell, shift = {(\Ox, \Oy)}, line width=\linewidthact ] 
        plot ({\x},{sin(1 * \x r)});
    \draw[-, scale=\scale,domain=-3:3,smooth,variable=\x,or, shift = {(\Ox, \Oy)}, line width=\linewidthact ] 
        plot ({\x},{sin(2 * \x r)});
    \draw[-, scale=\scale,domain=-3:3,smooth,variable=\x,americanrose, shift = {(\Ox, \Oy)}, line width=\linewidthact ] 
        plot ({\x},{sin(3 * \x r)});
    
    \draw[scale=\scale, shift = {(\Ox+8, \Oy)}] (-3,0) -- (4,0) node[right] {};
    \draw[scale=\scale, shift = {(\Ox+8, \Oy)}] (0,-1) -- (0,3) node[above] {SIREN \cite{sitzmann2019siren}};
    \draw[-, scale=\scale,domain=-3:3,smooth,variable=\x,yell, shift = {(\Ox+8, \Oy)}, line width=\linewidthact ] 
        plot ({\x},{sin(1 * \x r)});
    \draw[-, scale=\scale,domain=-3:3,smooth,variable=\x,or, shift = {(\Ox+8, \Oy)}, line width=\linewidthact ] 
        plot ({\x},{sin(2 * \x r)});
    \draw[-, scale=\scale,domain=-3:3,smooth,variable=\x,americanrose, shift = {(\Ox+8, \Oy)}, line width=\linewidthact ] 
        plot ({\x},{sin(3 * \x r)});
    
    \draw[scale=\scale, shift = {(\Ox+16, \Oy)}] (-3,0) -- (4,0) node[right] {};
    \draw[scale=\scale, shift = {(\Ox+16, \Oy)}] (0,-1) -- (0,3) node[above] {};
    \draw[-, scale=\scale,domain=-3:3,smooth,variable=\x,yell, shift = {(\Ox+16, \Oy)}, line width=\linewidthact ] 
        plot ({\x},{sin(1 * \x r)});
    \draw[-, scale=\scale,domain=-3:3,smooth,variable=\x,or, shift = {(\Ox+16, \Oy)}, line width=\linewidthact ] 
        plot ({\x},{sin(2 * \x r)});
    \draw[-, scale=\scale,domain=-3:3,smooth,variable=\x,americanrose, shift = {(\Ox+16, \Oy)}, line width=\linewidthact ] 
        plot ({\x},{sin(3 * \x r)});
    

    \draw[scale=\scale, shift = {(\Ox+28.5, \Oy-3)}] (-2,0) -- (3,0) node[right] {};
    \draw[scale=\scale, shift = {(\Ox+28.5, \Oy-3)}, line width=\linewidthact] (0,-1) -- (0,3) node[above] {sigmoid};
    \draw[-,scale=\scale,domain=-2:2,smooth,variable=\x,blue, shift = {(\Ox+28.5, \Oy-3)} ] plot ({\x},{1/(1+exp(-\x * 2))});
    
    \draw[scale=\scale, shift = {(\Ox+28.5, \Oy-16.5)}] (-1,0) -- (3,0) node[right] {};
    \draw[scale=\scale, shift = {(\Ox+28.5, \Oy-16.5)}] (0,-1) -- (0,3) node[above] {linear};
    \draw[-, scale=\scale,domain=-1:2,smooth,variable=\x,green, shift = {(\Ox+28.5, \Oy-16.5)}, line width=\linewidthact ] plot ({\x},{(\x)});
    
\end{tikzpicture}


}
    \hfill
    \resizebox{0.32\textwidth}{!}{
\def\layersep{2cm}
\def\nodesep{0.3cm}
\def\linewidthact{0.18mm}
\definecolor{americanrose}{rgb}{1.0,0.01,0.24}
\definecolor{or}{rgb}{1.0,0.49,0.0}
\definecolor{yell}{rgb}{1.0,0.75,0.0}

\newcommand{\randomcolor}{
    \pgfmathsetmacro{\randnum}{int(random(1,3))}
    \ifnum\randnum=1
        \def\colorselect{americanrose}
    \else\ifnum\randnum=2
        \def\colorselect{or}
    \else
        \def\colorselect{yell}
    \fi\fi
}
\newcommand{\hiddenneuronrandomcolor}{
    \randomcolor
    \tikzstyle{hidden neuron softplus}=[neuron_hidden, fill=\colorselect]
}

\begin{tikzpicture}[shorten >=1pt,->,draw=black!50, node distance=\layersep, scale=0.7]
    \tikzstyle{every pin edge}=[<-,shorten <=1pt,line width=0.5pt,color=black!85, opacity=0.5]

    \tikzset{mystyle/.style={line width=0.025mm,color=black!75, opacity=0.5}}

    \tikzstyle{neuron}=[circle,fill=black!25,minimum size=3pt,inner sep=0pt]

    \tikzstyle{neuron_hidden}=[circle,fill=black!25,minimum size=3pt,inner sep=0pt]
    
    \tikzstyle{input neuron}=[neuron, fill=gray];
    \tikzstyle{output neuron}=[neuron, fill=blue];
    \tikzstyle{output neuron2}=[neuron, fill=blue];

    \tikzstyle{hidden neuron softplus}=[neuron_hidden, fill=blue!50];

    \tikzstyle{annot} = [text width=4em, text centered]

    \node[input neuron, pin=left:$\mathbf{p}_3$] (I-1) at (0,-\nodesep*1*2.2) {};
    \node[input neuron, pin=left:$\mathbf{v}_3$] (I-2) at (0,-\nodesep*2*2.2) {};
    \node[input neuron, pin=left:$\boldsymbol{\lambda}_3$] (I-3) at (0,-\nodesep*3*2.2) {};
    \node[input neuron, pin=left:$\boldsymbol{\Omega}_3$] (I-4) at (0,-\nodesep*4*2.2) {};
    \node[input neuron, pin=left:$\boldsymbol{\omega}_4$] (I-5) at (0,-\nodesep*5*2.2) {};

    \foreach \name / \y in {1,...,32}
        \hiddenneuronrandomcolor
        \path[yshift=0.5cm]
            node[hidden neuron softplus] (H1-\name) at (\layersep,-\nodesep*0.5*\y) {};

    \foreach \name / \y in {1,...,32}
        \hiddenneuronrandomcolor
        \path[yshift=0.5cm]
            node[hidden neuron softplus] (H2-\name) at (1.8*\layersep,-\nodesep*0.5*\y) {};
            
    \foreach \name / \y in {1,...,32}
        \hiddenneuronrandomcolor
        \path[yshift=0.5cm]
            node[hidden neuron softplus] (H3-\name) at (2.6*\layersep,-\nodesep*0.5*\y) {};

    \node[output neuron2, pin={[pin edge={->}]right:$\omega_1$}, right of=H3-3] (O-1) at (2.3*\layersep,-1.0)  {};
    \node[output neuron2, pin={[pin edge={->}]right:$\omega_2$}, right of=H3-3] (O-2) at (2.3*\layersep,-1.6)  {};
    \node[output neuron2, pin={[pin edge={->}]right:$\omega_3$}, right of=H3-3] (O-3) at (2.3*\layersep,-2.2)  {};
    \node[output neuron2, pin={[pin edge={->}]right:$\omega_4$}, right of=H3-3] (O-4) at (2.3*\layersep,-2.8)  {};

    \pgfdeclarelayer{background}
    \pgfsetlayers{background,main}

    \begin{pgfonlayer}{background}

    \foreach \source in {1,...,5}
        \foreach \dest in {1,...,32}
            \path[mystyle] (I-\source) edge (H1-\dest);
            
    \foreach \source in {1,...,32}
        \foreach \dest in {1,...,32}
            \path[mystyle] (H1-\source) edge (H2-\dest);


    \foreach \source in {1,...,32}
        \foreach \dest in {1,...,32}
            \path[mystyle] (H2-\source) edge (H3-\dest);

    \foreach \source in {1,...,32}
        \foreach \dest in {1,...,4}
            \path[mystyle] (H3-\source) edge (O-\dest);

    \end{pgfonlayer}

    \pgfmathsetmacro{\scale}{0.2}
    \pgfmathsetmacro{\Ox}{10}
    \pgfmathsetmacro{\Oy}{4}

    \draw[scale=\scale, shift = {(\Ox, \Oy)}] (-3,0) -- (4,0) node[right] {};
    \draw[scale=\scale, shift = {(\Ox, \Oy)}] (0,-1) -- (0,3) node[above] {};
    \draw[-, scale=\scale,domain=-3:3,smooth,variable=\x,yell, shift = {(\Ox, \Oy)} ] 
        plot ({\x},{sin(1 * \x r)});
    \draw[-, scale=\scale,domain=-3:3,smooth,variable=\x,or, shift = {(\Ox, \Oy)} ] 
        plot ({\x},{sin(2 * \x r)});
    \draw[-, scale=\scale,domain=-3:3,smooth,variable=\x,americanrose, shift = {(\Ox, \Oy)}, line width=\linewidthact ] 
        plot ({\x},{sin(3 * \x r)});
    
    \draw[scale=\scale, shift = {(\Ox+8, \Oy)}] (-3,0) -- (4,0) node[right] {};
    \draw[scale=\scale, shift = {(\Ox+8, \Oy)}] (0,-1) -- (0,3) node[above] {SIREN \cite{sitzmann2019siren}};
    \draw[-, scale=\scale,domain=-3:3,smooth,variable=\x,yell, shift = {(\Ox+8, \Oy)}, line width=\linewidthact ] 
        plot ({\x},{sin(1 * \x r)});
    \draw[-, scale=\scale,domain=-3:3,smooth,variable=\x,or, shift = {(\Ox+8, \Oy)}, line width=\linewidthact ] 
        plot ({\x},{sin(2 * \x r)});
    \draw[-, scale=\scale,domain=-3:3,smooth,variable=\x,americanrose, shift = {(\Ox+8, \Oy)}, line width=\linewidthact ] 
        plot ({\x},{sin(3 * \x r)});
    
    \draw[scale=\scale, shift = {(\Ox+16, \Oy)}] (-3,0) -- (4,0) node[right] {};
    \draw[scale=\scale, shift = {(\Ox+16, \Oy)}] (0,-1) -- (0,3) node[above] {};
    \draw[-, scale=\scale,domain=-3:3,smooth,variable=\x,yell, shift = {(\Ox+16, \Oy)}, line width=\linewidthact ] 
        plot ({\x},{sin(1 * \x r)});
    \draw[-, scale=\scale,domain=-3:3,smooth,variable=\x,or, shift = {(\Ox+16, \Oy)}, line width=\linewidthact ] 
        plot ({\x},{sin(2 * \x r)});
    \draw[-, scale=\scale,domain=-3:3,smooth,variable=\x,americanrose, shift = {(\Ox+16, \Oy)}, line width=\linewidthact ] 
        plot ({\x},{sin(3 * \x r)});
    
    \draw[scale=\scale, shift = {(\Ox+27.25, \Oy-6.5)}] (-2,0) -- (3,0) node[right] {};
    \draw[scale=\scale, shift = {(\Ox+27.25, \Oy-6.5)}] (0,-1) -- (0,3) node[above] {sigmoid};
    \draw[-,scale=\scale,domain=-2:2,smooth,variable=\x,blue, shift = {(\Ox+27.25, \Oy-6.5)}, line width=\linewidthact ] plot ({\x},{1/(1+exp(-\x * 2))});
    
    
\end{tikzpicture}


}
  \caption{From left to right: full G\&CNET architectures for interplanetary transfer, asteroid landing and drone racing.}
  \label{fig:ffnns}
\end{figure*}
Each G\&CNET has two hidden layers with 32 neurons each, which amounts to 2435, 2500 and 2788 parameters for the interplanetary transfer, the asteroid landing and the drone racing case, respectively. We train the G\&CNETs on datasets of optimal trajectories using behavioral cloning. Since this is a simple regression task, we also opt here for the SIREN architecture \cite{sitzmann2019siren}. In fact previous work already shows the impressive performance of SIREN for the same optimal control problems \cite{origer2024gcnetsiren}.
The full G\&CNET architectures are shown in Fig.\ref{fig:ffnns}.
For the optimal control problems which we solve using Pontryagin's Maximum principle (interplanetary transfer and asteroid landing), we leverage a technique called the "Backward Generation of Optimal Examples" (BGOE) \cite{izzo2021real,dario_seb_gcnet}. This allows us to generate very efficiently hundreds of thousands of optimal trajectories by perturbing one single nominal solution (400,000 optimal trajectories for the interplanetary transfer and 300,000 for the asteroid landing). In the drone racing case we solve 10,000 optimal trajectories individually using a direct method, see \cite{origer2023guidance}. The initial and final conditions for the optimal trajectories are listed in App.\ref{app:0}. Once these trajectories are obtained they are sampled in 100 points (interplanetary transfer and asteroid landing) and 199 points (drone racing). All these state-action pairs can then be used as features and labels respectively in the behavioral cloning pipeline.
In all the cases we use a 80/20 split for training and validation data, the Adam optimizer \cite{kingma2014adam} and a scheduler that decreases the learning rate by 10\% whenever the loss fails to improve for 10 consecutive epochs.
The loss function for the interplanetary transfer is: $\mathcal L= 1-\frac{\mathbf{\hat{t}}^*\cdot \mathbf{\hat{t}}_{nn}}{\mathbf{\vert\hat{t}}^*\vert\vert \mathbf{\hat{t}}_{nn}\vert}$, hence the G\&CNET learns to minimize the cosine similarity between the estimated thrust direction $\mathbf{\hat{t}}_{nn}$ and the ground truth $\mathbf{\hat{t}}^*$. For the asteroid landing we add an additional term which penalizes the mean squared error between the estimated throttle $u_{nn}$ and the ground truth $u^*$:  $\mathcal L= \text{MSE}(u_{nn},u^*) + 1-\frac{\mathbf{\hat{t}}^*\cdot \mathbf{\hat{t}}_{nn}}{\mathbf{\vert\hat{t}}^*\vert\vert \mathbf{\hat{t}}_{nn}\vert}$.
In the drone racing case, the loss function is simply the mean squared error between the estimated control input and the ground truth: $\mathcal \mathcal{L}=\text{MSE}(u_{nn},u^*)$.
For the interplanetary transfer, asteroid landing, and drone racing cases, we use the following hyperparameters: 4096 as the batch size (training and validation), $5\text{e-5}$ as the learning rate, weight decay values of $2.5\text{e-5}$, $2.5\text{e-5}$, and $0.0$ respectively, and training epochs of 500, 500, and 200 respectively.
The training and validation loss during training are depicted in Fig.\ref{fig:losses_figs} for the three optimal control problems.
\begin{figure*}[!t]
  \centering
  \includegraphics[width=0.32\textwidth]{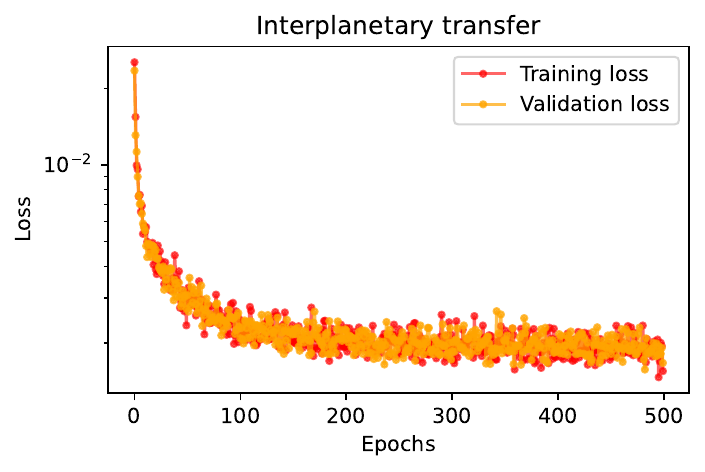}
  \hfill
  \includegraphics[width=0.32\textwidth]{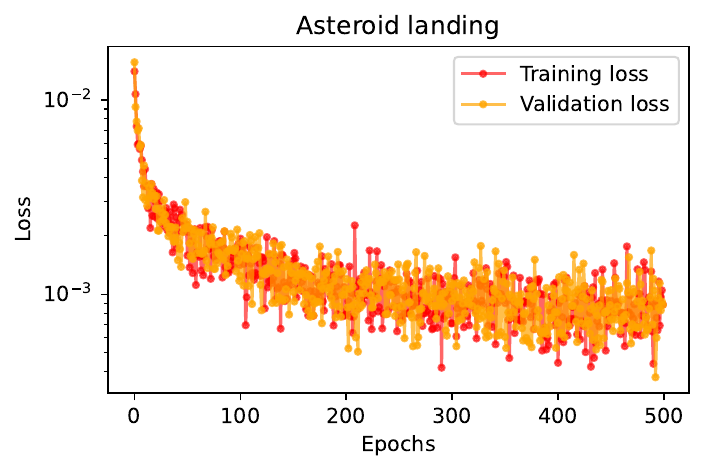}
  \hfill
  \includegraphics[width=0.32\textwidth]{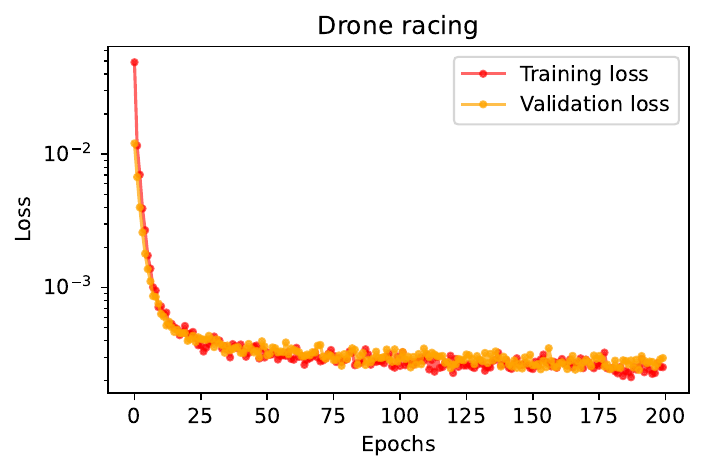}
  \hfill
  \caption{Training and validation loss of G\&CNETs.}
  \label{fig:losses_figs}
\end{figure*}

\section{Results \& Discussion}
\begin{figure*}[h!]
  \centering
  \includegraphics[width=\textwidth]{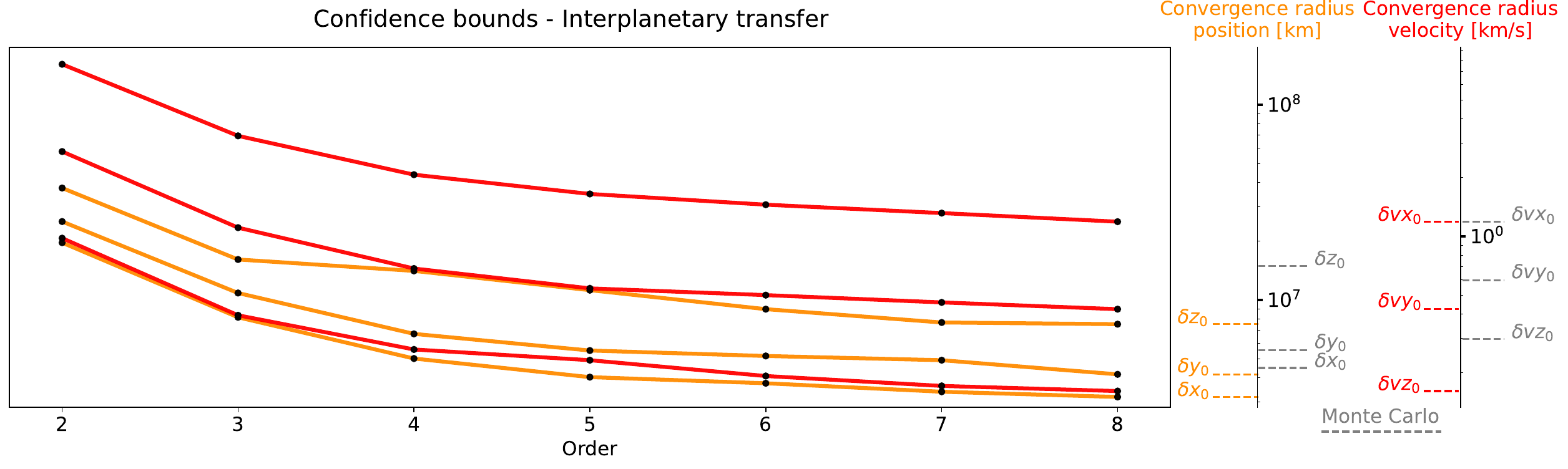}
  \hfill
  \includegraphics[width=\textwidth]{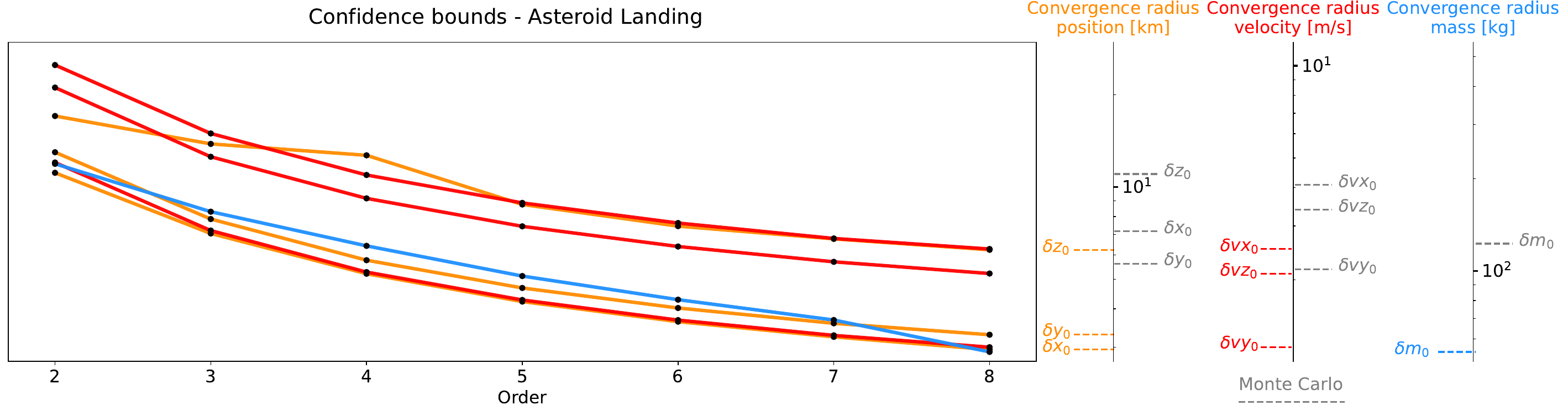}
  \hfill
  \includegraphics[width=\textwidth]{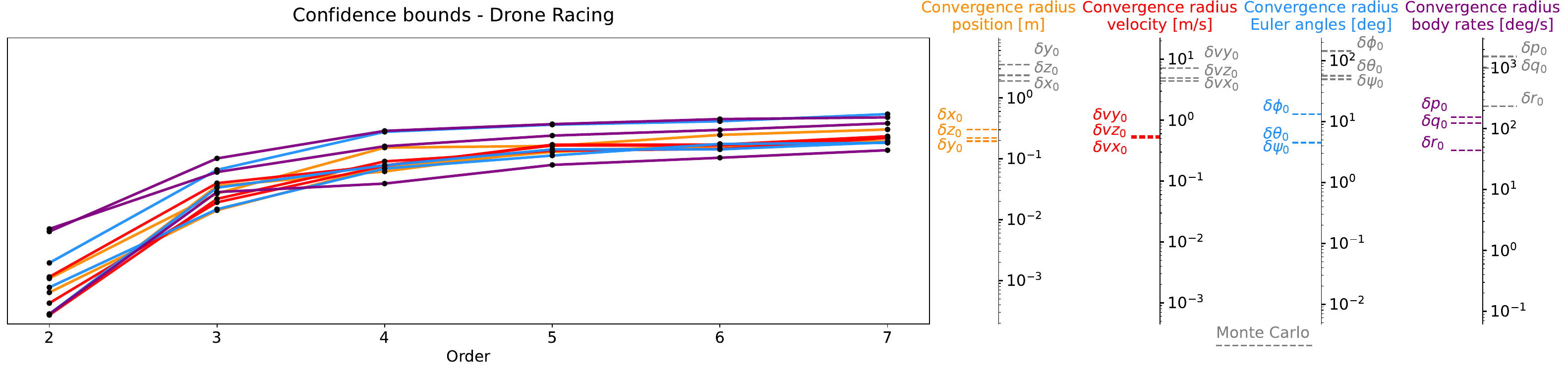}
  \caption{Proxy for convergence radius of each individual state as a function of polynomial order and Monte Carlo analysis.}
  \label{fig:cauchy}
\end{figure*}

The confidence bounds for each individual state of the three optimal control problems are shown in Fig.~\ref{fig:cauchy}. Each of these bounds corresponds to the radius of convergence of the polynomial provided the initial conditions are perturbed within that radius (from the nominal trajectory). For example, in the case of the interplanetary transfer, we can perturb the initial position along the z-axis by $\pm \delta z_0 \approx 7,500,000 $ km and still confidently use the polynomial to perform uncertainty analysis. There are two important caveats here. First, these are bounds on single states and not on multiple states simultaneously. Second, the bounds approach the true convergence radius as the polynomial order $\to \infty$ (Cauchy-Hadamard theorem). 
Therefore, the bounds we provide are approximations, as we achieve only 8th-order accuracy for the two first problems and 7th-order accuracy for the drone racing problem. Practically, one is limited to the computationally resources available. The memory requirements to store the variational integrator grow quickly with increasing order. With this, also the computational time required to assemble the variational equations and propagate all the equations.

As a comparison we also add the results from a Monte Carlo analysis of the G\&CNETs performance in Fig.~\ref{fig:cauchy} (dashed grey lines). 
We checked, for each state individually, that for 10,000 random perturbations of that state (where the dashed grey line represent the maximal perturbation), the G\&CNET still steers the system successfully to the target. In the case of the interplanetary transfer this meant reaching the sphere of influence of the target planet, for the asteroid landing it meant reaching $1$ km altitude above the asteroid surface with a relative velocity $\leq 15$ m/s (we added the velocity constraint because even for perturbation which the G\&CNET cannot handle, the trajectories eventually hit the asteroid sub-optimally), and in the drone racing case we checked that all the trajectories passed through the gate. 
Assuming the polynomials are \textit{analytic} (this requires the event manifold to be analytic in the first place), then the series are point-wise convergent. This means that, within the radius of convergence, the Taylor series converges to the function it approximates. Hence, the radius of convergence of the series will always be smaller than the maximal perturbation which the G\&CNET can handle, past which it misses the event manifold.
Indeed, the maximal allowed perturbations predicted by the Monte Carlo analysis are always larger than the ones provided by the Cauchy-Hadamard theorem, see Fig.~\ref{fig:cauchy}.

As a side note, the accuracy of G\&CNETs has improved considerably over the last years while the amount of parameters used to train the policies decreased substantially \cite{izzo2021real, dario_seb_gcnet, origer2024closing, origer2024gcnetsiren}. For this interplanetary transfer, initial state uncertainties of $\pm 3\text{e}{6}$ km in position and $\pm 0.3$ km/s in velocity should be well within real mission requirements. Similarly, for the asteroid landing the G\&CNET can handle initial state uncertainties of $\pm 3$ km in position, $\pm 1$ m/s in velocity and $\pm 65$ kg in mass.

\begin{figure}[h!]
   \centering
\includegraphics[width=\columnwidth]{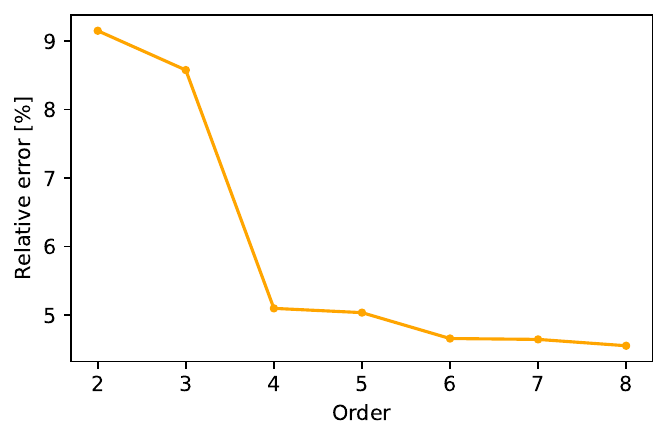}
\caption{Relative error (Frobenius norm) between Monte Carlo- and Moment-Generating-Function -based covariance matrices versus polynomial order.}\label{fig:rel_error_mc_mgf}
\end{figure}

Let's go back to the requirement posed in the introduction and see how we can address it:
\begin{quote}
The G\&C algorithm shall steer the spacecraft to an altitude of $1$ km $\pm 5$ m above the asteroid surface and ensure a relative velocity of $\leq 15$ m/s. The algorithm must achieve this relative velocity within the specified limit in at least 95\% of scenarios, given an initial state uncertainty of $\pm 5\%$ in spacecraft mass.
\end{quote}
The uncertainty in altitude $1$ km $\pm 5$ captures the accuracy of the neural network which learned to implicitely represent the event manifold.
We assume that the initial state uncertainty of 5\% in initial spacecraft mass is uniformly distributed in $m_0 \in [0.95m_0, 1.05m_0] \approx [335, 370]$ kg (here a different type of distribution could be chosen too~\cite{acciarini2024mgf}). Note that these bounds are well within the radius of convergence provided by Cauchy-Hadamard (see Fig.\ref{fig:cauchy}), hence we can confidently use this polynomial to perform uncertainty propagation. We compute the first two moments using moment-generating functions. The mean $\boldsymbol{\mu} = [\mu_{v_x}, \mu_{v_y}, \mu_{v_z}]$ and the covariance matrix $\boldsymbol{\Sigma} $ of the velocity components turn out to be:
\[
\boldsymbol{\mu} \approx
\begin{bmatrix}
    -7.98 \\
    6.80 \\
    -0.42
\end{bmatrix} \text{m/s}\, ,
\quad 
\boldsymbol{\Sigma} \approx
\begin{bmatrix}
    2.46 & -1.91 & 0.03 \\
    -1.91 & 1.48 & -0.03\\
    0.03& -0.03 & 0.001
\end{bmatrix} (\text{m/s})^2\, .
\]
Assuming that the velocity components are normally distributed, we can draw samples from the corresponding multivariate normal distribution and check the proportion of trajectories which satisfy the requirement on the velocity magnitude. In this case, the G\&CNET brings the spacecraft to the required altitude with a velocity magnitude $\leq 15$ m/s in $98\%$ of cases.
Of course more advanced statistical tools could be used at this stage and one should make use of the fact the higher statistical moments can be computed with moment-generating functions. In addition, such functions might be used in the future to inform the training process of G\&CNETs, for instance to increase their robustness to state uncertainties.
Finally, we verify our methodology by showing how the Frobenius norm between the complete covariance matrix obtained through MC ($20,000$ samples) and MGFs change as the polynomial order increases in Fig.~\ref{fig:rel_error_mc_mgf}.

\section{Conclusion}
Uncertainty propagation onto an event manifold via Differential Algebra is presented in the context of Guidance \& Control Networks (G\&CNETs). 
Three optimal control problems with event manifolds of varying complexity have been considered: a time-optimal interplanetary transfer, a mass-optimal asteroid landing, and energy-optimal drone racing. 
In all cases, we perform a high-order expansion of the final states on the event manifold as a function of the initial conditions and factor out the dependence on time by inverting the polynomials.
We subsequently provide confidence bounds using the Cauchy-Hadamard theorem, allowing us to confidently use the polynomials so long as the initial conditions are perturbed within the provided radii of convergence. 
Additionally, we apply Taylor methods in combination with moment-generating functions to compute the statistical moments of the final states given initial state uncertainties.
The provided methodology has the advantage that the robustness of G\&CNETs can be studied at a specific stage of the mission defined by an event manifold, as opposed to being limited to the study of locally stable points.
Finally, this work is driven by the recognition that there is a need to increase confidence in neural network certification for future missions.













\begin{appendices}

\section{Drone racing model}\label{app:1}
\begin{figure}[h!]
  \centering
  \includegraphics[width=0.6\columnwidth]{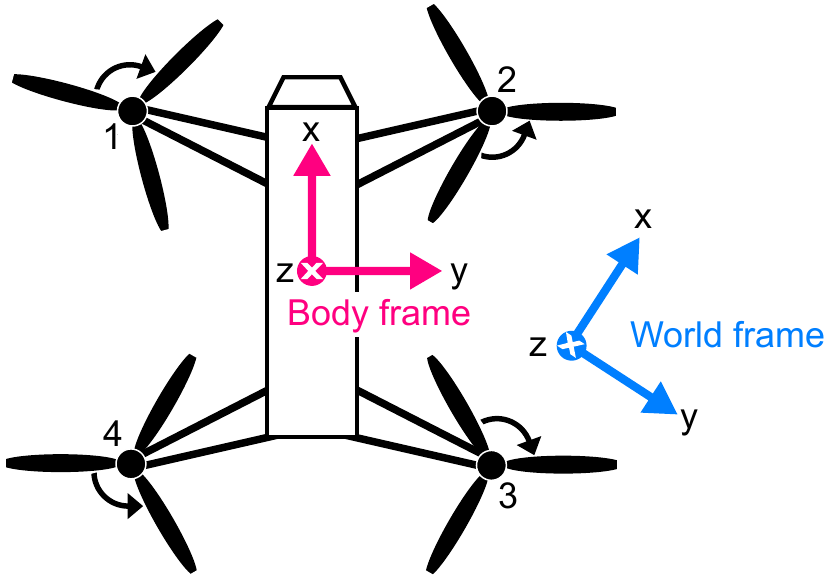}
  \caption{Coordinate frames (Body x-axis points to the front of the drone) \cite{origer2023guidance}.\vspace{-0mm}}
  \label{fig:coordframes}
\end{figure}

The coordinate frames used for the drone are shown in Fig.\ref{fig:coordframes}.
The rotation matrix \( R(\mb{\lambda}) \) converts coordinates from the body frame to the world frame. The notation \( c_{\theta} \) and \( s_{\phi} \) represents the cosine and sine of the respective Euler angles, respectively:
\begin{equation*}\label{eq:Euler_To World}
R(\mb{\lambda}) = 
\begingroup 
\setlength\arraycolsep{2pt}
\begin{bmatrix}
        c_{\theta}c_{\psi} & -c_{\phi}s_{\psi}+s_{\phi}s_{\theta}c_{\psi} & s_{\phi}s_{\psi}+c_{\phi}s_{\theta}c_{\psi} \\
        c_{\theta}s_{\psi} & c_{\phi}c_{\psi}+s_{\phi}s_{\theta}s_{\psi} & -s_{\phi}c_{\psi}+c_{\phi}s_{\theta}s_{\psi} \\
        -s_{\theta} & s_{\phi}c_{\theta} & c_{\phi}c_{\theta}
\end{bmatrix}
\endgroup
\end{equation*}
and $Q(\mb{\lambda})$ is the inverse transformation matrix: 
\begin{equation*} \label{eq:WorldToEuler}
Q(\mb{\lambda}) = 
\begin{bmatrix}
    1 & \sin{\phi} \tan{\theta}     & \cos{\phi} \tan{\theta} \\
    0 & \cos{\phi}                  & -\sin{\phi} \\
    0 & \sin{\phi} / \cos{\theta}   & \cos{\phi} / \cos{\theta}
\end{bmatrix}\, .
\end{equation*}
Note that the superscript \(\square^B\) indicates the body frame, and all model parameters are sourced from \cite{origer2023guidance}.
The forces \(\mb{F} = [F_x, F_y, F_z]^T\) are calculated based on the thrust and drag model from \cite{thrust_and_drag_model}: 
\begin{align} \label{eq:F_model}
\begin{split}
    F_x &= - k_x v^B_x \sum_{i=1}^4 \omega_i \,\, ,\quad
    F_y = - k_y v^B_y \sum_{i=1}^4 \omega_i\\
    F_z &= -k_\omega \sum_{i=1}^4 \omega_i^2 - k_z v^B_z \sum_{i=1}^4 \omega_i - k_h (v^{B2}_x + v^{B2}_y)\, .
\end{split} 
\end{align}
Finally the moment equations $\mb{M}=[M_x, M_y, M_z]^T$ are:
\begin{equation} \label{eq:M_model}
\begin{split}
    M_x &= k_p (\omega_1^2 - \omega_2^2 - \omega_3^2 + \omega_4^2) + k_{pv} v^{B}_y\\
    M_y &= k_q (\omega_1^2 + \omega_2^2 - \omega_3^2 - \omega_4^2) + k_{qv} v^{B}_x\\
    M_z &= k_{r1} (-\omega_1 + \omega_2 - \omega_3 + \omega_4) \\
    & + k_{r2} (-\dot{\omega}_1 + \dot{\omega}_2 - \dot{\omega}_3 + \dot{\omega}_4)  - k_{rr} r\\
\end{split}
\end{equation}

\section{Optimal control problems values}
The initial and final conditions for the optimal trajectories, as well as constants, are provided in Tab.\ref{tab:val1}.
\label{app:0}
\begin{table*}[!ht]
    \centering
    \caption{Initial conditions, final conditions and constants used in the optimal control problems. \label{tab:val1}}
    \begin{tabular}{c|cc|cc|c||cc}
    
Problem & \multicolumn{2}{c|}{Initial conditions}  & \multicolumn{2}{c|}{Final conditions}  & Frame & \multicolumn{2}{c}{Constants} 

\\\hline \hline

\multirow{5}{*}{Interplanetary} 
& $x_0$ & $-1.1874388$ AU & $x_f$ ($R$) & $1.3$ AU & \multirow{6}{*}{$\mathcal F$} &  $\Gamma$ & $0.1$ mm/s$^2$ \\ 
\multirow{5}{*}{transfer} & $y_0$ & $-3.0578396$ AU & $y_f$ & $0$ AU &  & $\mu$    &  (Sun) m$^3$/s$^2$\\ 
& $z_0$ & $0.3569406$ AU & $z_f$ & $0$ AU && \\ 
& $v_{x_0}$ & $-48.17$ km/s & $v_{x_f}$ & $0$ km/s && \\ 
& $v_{y_0}$ &  $18.30$ km/s & $v_{y_f}$ & $0$ km/s &&  \\ 
& $v_{z_0}$ &  $0.64$ km/s & $v_{z_f}$ & $0$ km/s  && \\ 

 \hline\hline
\multirow{6}{*}{Asteroid} 
& $x_0$ & $180$ km & $x_f$ & $122.2$ km & \multirow{6}{*}{$\mathcal R$} &    $I_{sp}$   &  $600$ s     \\ 
\multirow{6}{*}{landing}& $y_0$ & $-4.8$ km & $y_f$ & $90.33$ km & &  $g_0$   &   $9.8$ m/s$^2$  \\ 
& $z_0$ & $0$ km & $z_f$ & $-1.6$ km  & &   $c_1$   &  $80$ N    \\ 
& $v_{x_0}$ & $25$ m/s & $v_{x_f}$ & $0$ m/s & &  $\mu$    &    $1530348199$ m$^3$/s$^2$    \\ 
& $v_{y_0}$ &  $-25$ m/s & $v_{y_f}$ & $0$ m/s & &   $\omega$  &  $0.00041596$ rad/s   \\ 
& $v_{z_0}$ &  $20$ m/s & $v_{z_f}$ & $0$ m/s  &&      \\
\cline{2-6}
& $m_0$ &  $353$ kg & $m_f$ & left free  & -   &&    \\ 

 \hline\hline
\multirow{13}{*}{Drone} 
& $x_0$ & $\in [-5.0,-2.0]$ m & $x_f$ & $0$ m & \multirow{9}{*}{$\mathcal W$} &    $\omega_{min}$   &  $3000$ rpm     \\ 
\multirow{13}{*}{racing}& $y_0$ & $\in [-1.0,1.0]$ m & $y_f$ & $0$ m & &  $\omega_{max}$   &   $12000$ rpm  \\ 
& $z_0$ & $\in [-0.5,0.5]$ m & $z_f$ & $0$ m  & &    &     \\ 
& $v_{x_0}$ & $\in [-0.5,5.0]$ m/s & $v_{x_f}$ & left free & &     &        \\ 
& $v_{y_0}$ &  $\in [-3.0,3.0]$ m/s & $v_{y_f}$ & left free & &     &     \\ 
& $v_{z_0}$ &  $\in [-1.0,1.0]$ m/s & $v_{z_f}$ & left free  &&      \\
& $\phi_0$ & $\in [-40,40]$ deg & $\phi_f$ & $0$ deg & &     &        \\ 
& $\theta_0$ &  $\in [-40,40]$ deg & $\theta_f$ & $0$ deg & &     &     \\ 
& $\psi_0$ &  $\in [-60,60]$ deg & $\psi_f$ & $45$ deg  &&      \\
\cline{2-6}
& $p_0$ & $\in [-1,1]$ rad/s & $p_f$ & $0$ rad/s & \multirow{3}{*}{$\mathcal B$} &     &        \\ 
& $q_0$ &  $\in [-1,1]$ rad/s & $q_f$ & $0$ rad/s & &     &     \\ 
& $r_0$ &  $\in [-1,1]$ rad/s & $r_f$ & $0$ rad/s  &&      \\
\cline{2-6}
& $\boldsymbol{\omega}_0$ &  $7500$ rpm & $\boldsymbol{\omega}_f$ & left free  & -   &&    \\ 

    
    
 \hline\hline
    \end{tabular}
\end{table*}

\end{appendices}

\clearpage
\bibliography{biblio}


\end{document}